\documentclass[journal=ancac3,manuscript=article]{achemso}

\author{Alexis Devilez}
\author{Nicolas Bonod}
\email{nicolas.bonod@fresnel.fr}
\author{Brian Stout}
\affiliation{Institut Fresnel, Aix-Marseille Universit\'e, CNRS \\
Domaine Universitaire de Saint J\'er\^ome, 13397 Marseille, France}

\title[\texttt{achemso} demonstration]
{Compact metallo-dielectric optical antenna for ultra directional and enhanced radiative emission}

\begin{document}
%%%%%%%%%%%%%%%%%%%%%%%%%%%%%%%%%%%%%%%%%%%%%%%%%%%%%%%%%%%%%%%%%%%%%
%% The manuscript does not need to include \maketitle, which is
%% executed automatically.  The document should begin with an
%% abstract, if appropriate.  If one is given and should not be, the
%% contents will be gobbled.
%%%%%%%%%%%%%%%%%%%%%%%%%%%%%%%%%%%%%%%%%%%%%%%%%%%%%%%%%%%%%%%%%%%%%
\begin{abstract}
We report the design of highly efficient optical antennas employing a judicious synthesis of metallic and dielectric materials. In the proposed scheme, a pair of metallic coupled nanoparticles permits large enhancements in both excitation strength and radiative decay rates, while a high refractive index dielectric microsphere is employed to efficiently collect light without spoiling the emitter quantum efficiency. Our simulations indicate potential fluorescence rate enhancements of 3 orders of magnitude over the entire optical frequency range.
\end{abstract}

\section{Introduction}

Photoluminescence signal detection of fluorescent molecules or quantum dots is a crucial issue in many photonic applications. The recent concept of optical nanoantennas is based on using plasmonic nano-structures to tailor the electromagnetic environment near a quantum emitter in order to enhance the photoluminescence signal by optimizing : (a) the local excitation rate, (b) the emission rate, and (c) the collection efficiency.\cite{muhl_science05, Greff_scien,review_novotny}
During the past decade, several types of metallic nanostructures, in particular, coupled metallic particles \cite{Ag_nnpt_pairs, prop_gold_nnpt, stocknanolens,BidaultADN08} and subwavelength holes \cite{Rig_holes_PRL, popov_holes} have demonstrated their ability to confine light in nanometer scale volumes and strongly enhance the excitation rate of single emitters placed in their vicinity. 

Reciprocally, nanoscale metallic structuring modifies the local density of states\cite{Ldos_interf}, thereby modifying the radiation properties of nearby photoemitters. Metallic structures introduce electromagnetic relaxation channels such as plasmonic modes that can strongly enhance the total decay rate \cite{superemitter, coupled_elipse_Sando}. However, due to their  lossy nature, non radiative relaxation rates can become preponderant when an emitter is located too close to a metallic structure \cite{Greffet_single, Decay_dipole_carminati, Mertens_1nnp,  Sando_FDTD, Colas_fluo, lifetime_nanodimer}. This quenching effect can spoil the benefits of plasmon excitation. Consequently, it is crucially important in nano-antenna applications to determine the quantum efficiency, defined as the ratio between the radiative and total decay rates, particularly for small separations between the emitter and the metallic structure. Photoluminescence enhancement thus relies on a trade-off between excitation rate enhancements and high quantum efficiencies\cite{Quench_novotny}.
 
Additionally, recent papers have addressed the important issue of the angular redistribution of radiated power induced by the structured local environment. Recently, the Yagi-Uda radio antenna has been successfully downscaled to the optical range to obtain high directionality of light radiation \cite{YU_VHulst, YU_koenderink, YU_engheta}. The high directionality obtained with these structures requires couplings between the emitted light and the plasmon modes, necessitating a rather large number of nanoparticles. The size of a metallic director can be reduced to two coupled nanoparticles \cite{YU_engheta, Kallantenna} but at a price of a lower directivity. 
The design of an optical antenna based on a fully plasmonic approach is still challenging since high directionality can suffer from ohmic losses inherent to metallic structures and requires precise manufacturing techniques to align several particles.
Dielectric materials have usually been disregarded in this context because their extinction cross-sections are comparable in size to their geometrical cross-section, thus producing weak excitation rates. However, it has been demonstrated that high refractive index dielectric materials can induce a redirection of the dipolar emission \cite{Soller_interf, TiO2_fluo, Davy_collec}. Arrays of dielectric nanoparticles have also shown their ability to redirect light \cite{Tio2_array}. The considerable advantage of transparent (lossless) dielectrics is that they preserve the quantum efficiency.

This work demonstrates that a judicious combination of dielectric and metallic materials can produce highly directional compact optical antenna which strongly enhances both local field excitation and radiation rate of a dipolar emitter. We will see that desirable properties occur in conjunction with a large quantum efficiency thereby ensuring a high fluorescence emission rate.

In this article, numerical simulations are performed within the framework of rigorous Mie theory combined with a multiple scattering formulation\cite{stout08, FoldyLax, wiscombe}. The single emitter is simulated as a dipolar source modeled by taking the first electric term in a multipole (Mie) expansion. The generalized Lorentz-Mie calculations are performed with a truncation order of $n_{\rm max}$~ =~20 for the individual scatterers. The total emitted power $P_{\rm tot}$ and the radiative emitted power $P_{\rm rad}$ are calculated by integrating the radial component of the Poynting vector over a spherical surface surrounding the source at respective distances of 1~nm and 5~$\mu$m. We remark in passing that this partially numerical method for determining power emissions, validates more rapid quasi-analytical methods that we have also developed for calculating these quantities (to be presented elsewhere). The total and radiative decay rate enhancements are then obtained by normalizing the emitted power in the presence of the antenna by the emitted power in the homogeneous background medium: $\Gamma_{\rm tot}~=~P_{\rm tot}/P_0$ and $\Gamma_{\rm rad}~=~P_{\rm rad}/P_0$. The quantum efficiency is then defined as:
\begin{equation}
  \eta =\frac{\Gamma_{rad}}{(\Gamma_{tot}+(1-\eta_i)/\eta_i)} \label{eqn:effic}
\end{equation}
where $\eta_i$ represents the intrinsic quantum efficiency of the emitter 
($\eta_i~=~1$ for a perfect emitter). 

\begin{figure}[htbp]
\includegraphics[width=8cm]{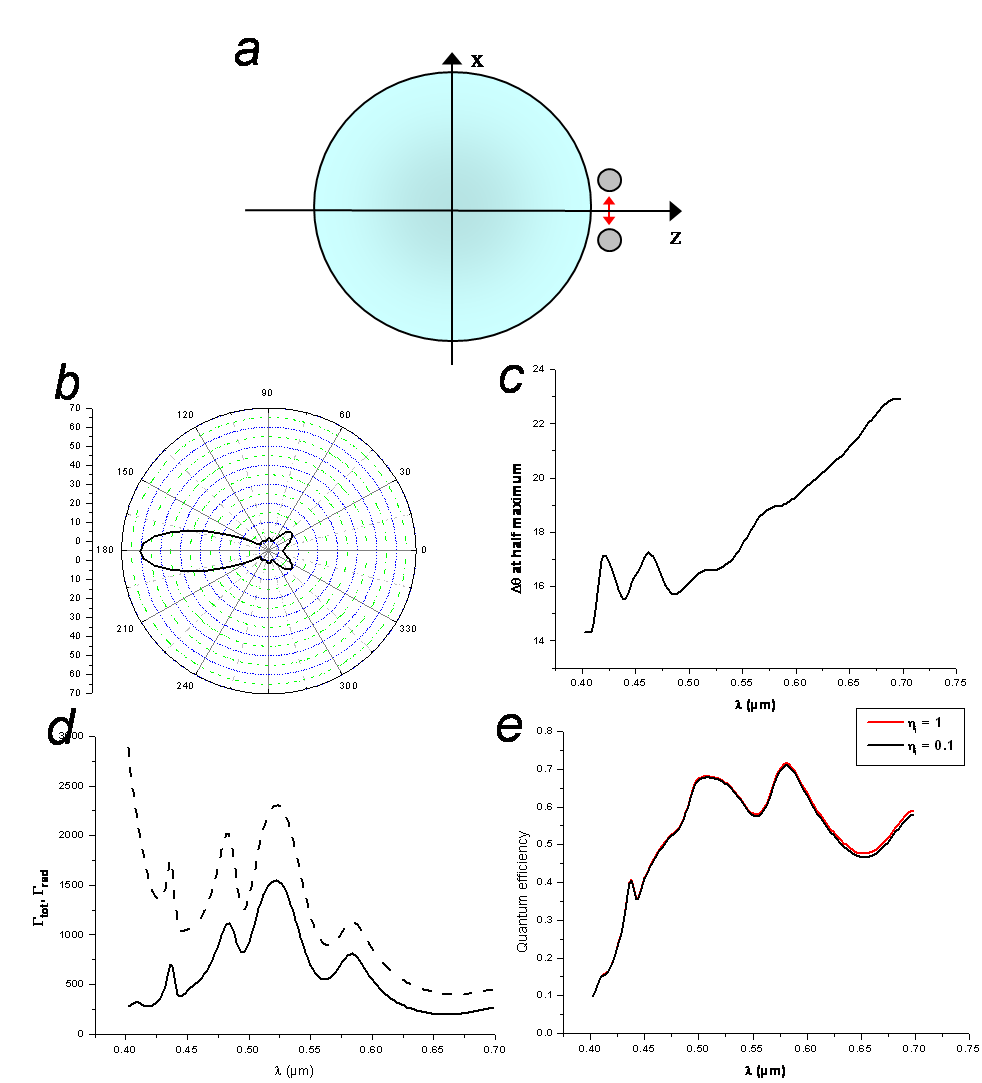}% Here is how to import EPS art
\caption{ a) Antenna schematic: a dielectric microsphere ($D$~=~500~nm, $n_{{\rm Tio}_2}$  taken from \citeauthor{rutile_Yamada}~\cite{rutile_Yamada}) and two silver nanoparticles separated by 8~nm ($D$~=~60~nm and $n_{\rm Ag}$ taken from Palik~\cite{Palik}) embedded in a dielectric background of refractive index $n_0$~=~1.3. A dipolar emitter is placed equidistantly from the silver spheres along their axis of separation. b) Radiation diagram at $\lambda~=~525$~nm, c) collection angle (3db half width) d) total (dashed line) and radiative (full line) decay rate enhancements and e)
 quantum efficiency for a perfect emitter (red line) and a  poor emitter, $\eta_i$~=~0.1 (black line).}
\label{fig:results}
\end{figure}

The schematic of the antenna being considered is displayed in \ref{fig:results}$a$. The optical antenna consists of a TiO$_2$ dielectric microsphere \cite{TiO2_fluo, rutile_Yamada} ($D$~=~500~nm) and two silver nanospheres ($D$~=~60~nm separated by 8~nm) embedded in a dielectric background of refractive index $n_0$~=~1.3. A dipolar-like source oriented along the $x$-axis is placed 30~nm from the dielectric microsphere and centered equidistantly along the axis joining the silver spheres.
\ref{fig:results}$b$ displays the radiation pattern of the antenna (polar plot of the radiated power per angle unit) for a dipole oscillating at $\lambda$~=~525~nm ($\lambda$ is the wavelength in vacuum). Contrary to free space radiation, \ref{fig:results}$b$ exhibits a highly directional emission with an angular aperture (3dB half-width) $\simeq 15^{\circ}$. Calculations displayed in \ref{fig:results}$c$ show that the collection angle remains below $25^{\circ}$ for a wavelength range from 400~nm to 700~nm.
More interestingly, this highly directive radiation is concurrent with a strong enhancement of the radiative decay rate. The radiative decay rate enhancement (full line) displayed in \ref{fig:results}$d$ surpasses two orders of magnitude over the entire optical frequency range. Moreover, the spectral feature of $\Gamma_{\rm rad}$ shows several peaks higher than $10^3$ which are attributed to electromagnetic structural resonances. In other words, this optical antenna does not require an optimization of the electromagnetic resonances in order to be efficient, but an optimization of the electromagnetic couplings between the dipolar emitter and electromagnetic resonances permits radiative decay factors $\Gamma_{\rm rad}$ as high as $10^3$. The high quality of the antenna is confirmed in \ref{fig:results}$e$ which shows a quantum efficiency greater than 0.5 for $\lambda~>~450$~nm. Even more spectacularly, we remark that the quantum efficiency of a poor emitter (black line in \ref{fig:results}$e$) is essentially the same as the quantum efficiency of a perfect emitter (red line). Further calculations show that this property is fulfilled for any intrinsic quantum efficiency higher than $10^{-3}$. This behavior is mainly due to the giant decay rates enhancements which render the term $(1-\eta_i)/\eta_i$ in \ref{eqn:effic} negligible compared to $\Gamma_{\rm tot}$.

In summary, this simple and compact system turns out to be a highly performant antenna over the entire optical spectrum and is characterized by a high directionality, strong decay rate enhancements, and a high quantum efficiency. In the remainder of this work, we investigate the properties of the dielectric and metallic components separately in order to better understand the unique performance of this optical antenna.
 
\begin{figure}[htbp]
\includegraphics[width=9cm]{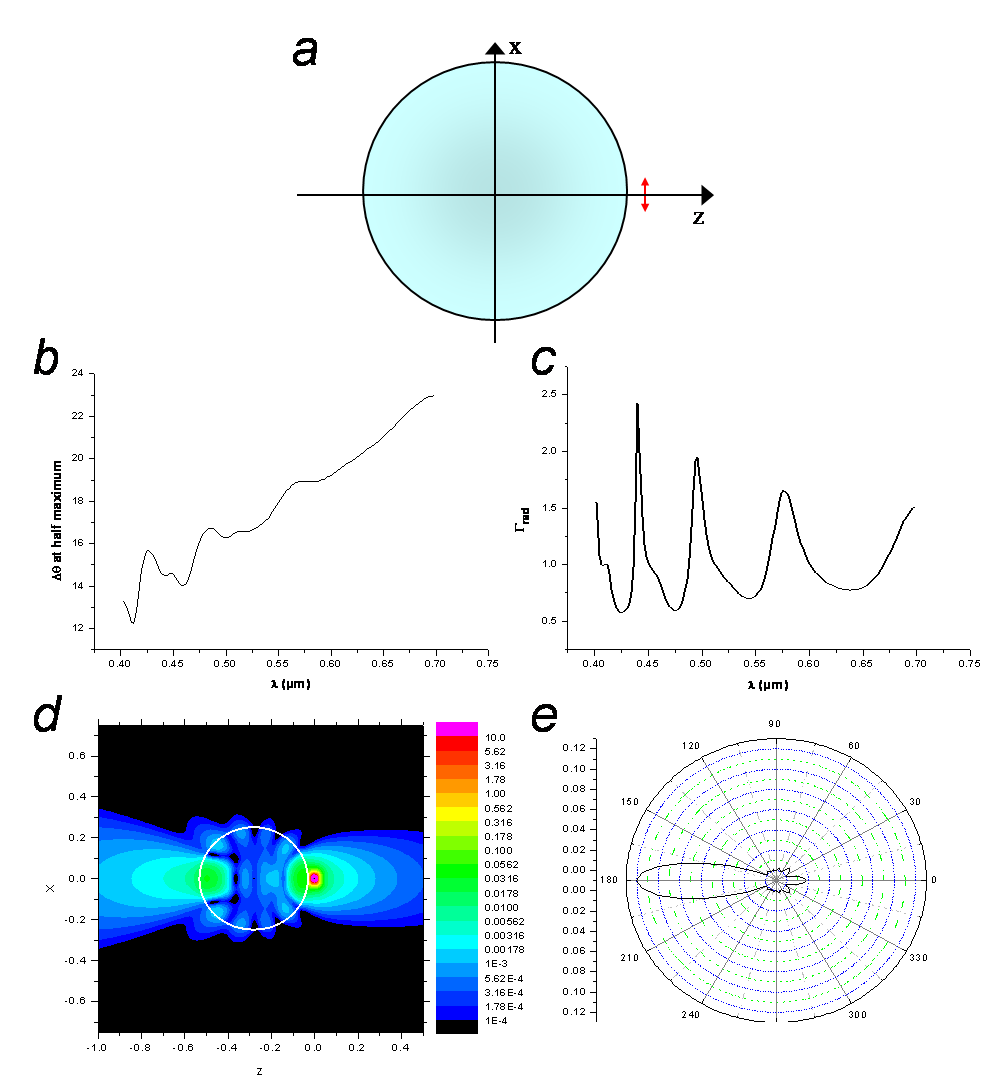}% Here is how to import EPS art
\caption{Characterization of TiO$_2$ microsphere as optical antenna. a) The microsphere is embedded in a background medium of refractive index index $n_0$~=~1.3. Refractive index of TiO$_2$ is taken from Ref.~\citenum{rutile_Yamada}. b) 3dB half width , c) radiative decay rate, d) quantum efficiency as a function of the wavelength and e) radiation diagram at $\lambda~=~440~nm$.}
\label{fig:Tio2}
\end{figure}

We first investigate the properties of a single high refractive index dielectric microsphere. Recent studies have demonstrated that dielectric microspheres operate as efficient near field ``lenses'' with performances comparable to the state of art of high numerical aperture microscopes such as immersion lenses \cite{Immerg_lens, TiO2_fluo, Davy_confin, 3Dconf, Davy_collec}. \ref{fig:Tio2}$b$ displays the angular 3dB half width of the emerging propagative beam produced by a dipole located at 30~nm from the sphere surface (the dipole is oriented along the $x$-axis). It clearly demonstrates that a single TiO$_2$ microsphere 500~nm in diameter can redirect dipolar radiated power into a narrow beam of 3dB half width below $25^{\circ}$ over a large frequency bandwidth covering almost all the optical range.
Furthermore, the radiative decay rate enhancement displayed in \ref{fig:Tio2}$c$  shows several peaks in the spectrum with moderated amplitudes. This behavior feature might seem surprising at first since dielectric materials generally exhibit low enhancement of the radiative decay rates. However, let us recall that the high refractive index of TiO$_2$ enables the well-known electromagnetic resonances in the dielectric microsphere called Whispering Gallery Modes (WGMs) \cite{WGM_theo}. This assertion is clearly demonstrated in \ref{fig:Tio2}$d$ which displays the electric field intensity in the vicinity of a  dielectric microsphere dipole illuminated at $\lambda~=~440$~nm corresponding to the narrowest peak in \ref{fig:Tio2}$c$. Furthermore, the far-field radiation pattern of emitted light at $\lambda~=~440$~nm (displayed in \ref{fig:Tio2}$e$) demonstrates that the excitation of WGMs does not significantly deteriorate the strong directional properties of the TiO$_2$ microsphere illuminated by a dipole. The WGM resonances do however have the ability to slightly enhance the radiative decay rates of a dipolar emitter. 
In summary, high refractive dielectric spheres can serve as simple and compact optical antennas in a very wide range of frequencies. Nevertheless, the huge decay rates enhancements observed in \ref{fig:results}$d$ cannot be explained by the coupling of a dipolar emitter with WGMs resonances. One concludes therefore that the bulk of the decay rate enhancements of the metallo-dielectric antenna is due to the metallic materials in the near field of the emitting dipole.

\begin{figure}[htbp]
\includegraphics[width=9cm]{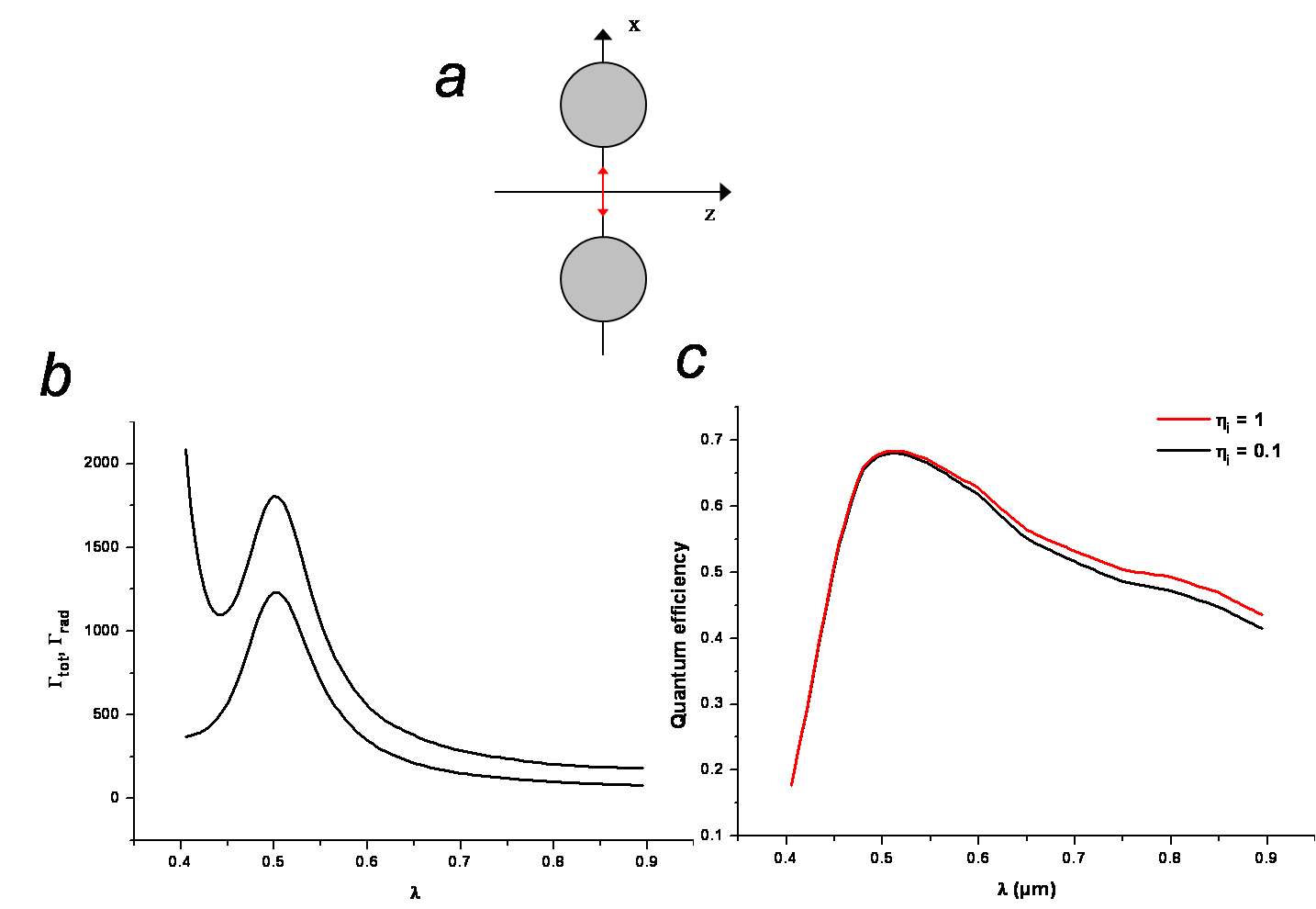}% Here is how to import EPS art
\caption{Characterization of the two silver nanoparticles ($D$~=~60~nm) embedded in a dielectric background of refractive index $n_0$~=~1.3. The dipolar light source is set midway between the two metallic elements separated by 8~nm and is assumed to be oriented along the separation axis. Refractive index of silver is taken from Ref.~\citenum{Palik}. b) Total (dashed line) and radiative (full line) decay rate enhancements, c) quantum efficiency for a perfect emitter (red line) and a poor emitter, $\eta_i$~=~0.1 (black line) as a function of the wavelength.}
\label{fig:superemitter}
\end{figure}

Coupled nanoparticles have been widely studied in the context of nanodimers \cite{dimer_prl08, lifetime_nanodimer} and bowtie nanoantennas  \cite{Bowtie_anten, coupled_elipse_Sando} to enhance the radiative decay rate of a single emitter. \ref{fig:superemitter}$b$ displays the total (dashed line) and radiative (full line) decay rate enhancements when a dipole oriented along the $x$-axis is set in the center of the cavity formed by two silver nanoparticles 60~nm in diameter separated by 8~nm. A broad peak appears corresponding to the well-known red-shifted plasmon resonance of two coupled metallic particles \cite{prop_gold_nnpt}. Comparison between \ref{fig:results}$c$ and \ref{fig:Tio2}$b$ shows that the so-called superemitter \cite{superemitter} does not significantly modify the radiation directionality when it is combined with a microsphere. As illustrated in \ref{fig:superemitter}$c$, the presence of metallic losses induces a drop of the quantum efficiency of a perfect emitter (red line), particularly strong near ultraviolet frequencies. Consequently, although the efficiency of a perfect emitter is decreased due to relaxation via non radiative channels, the quantum efficiency of a poor emitter is increased several-fold (c. f. \ref{fig:superemitter}$c$ with $\eta_i$~=~0.1).

The design of optical antennas has been widely inspired by their analogs in the radiofrequency range, the Yagi-Uda antenna in particular \cite{YU_engheta, YU_VHulst, YU_koenderink}. This antenna is typically made of three elements: the feed, the collector and the reflector. The feed element role is to improve couplings between the emitter and the antenna. The optical analogue of the collector generally consists of several coupled metal particles acting as a guide for plasmon waves, while the reflector can be made from a slightly larger single metallic particle. In our proposed metallo-dielectric antenna, the two coupled metallic particles act as the feed element, and the chain of guiding metallic particles and reflector are simply replaced by a single dielectric sphere. The lack of a reflector in the proposed metallo-dielectric antenna seem surprising, but it should be mentioned that optical reflector elements so far have not provided clear benefits either in the decay rate enhancements nor in the emission directionality. 
It can be shown that due to the dipole orientation, the addition of a reflector (made of a single metallic particle) decreases the radiative decay rates\cite{Mertens_1nnp}. Moreover, as illustrated in \ref{fig:Tio2}, the dielectric collector element is sufficiently performant to channel most of the emitted power without the need for an additional element contributing to the compactness of the antenna.

\begin{figure}[htbp]
\includegraphics[width=9cm]{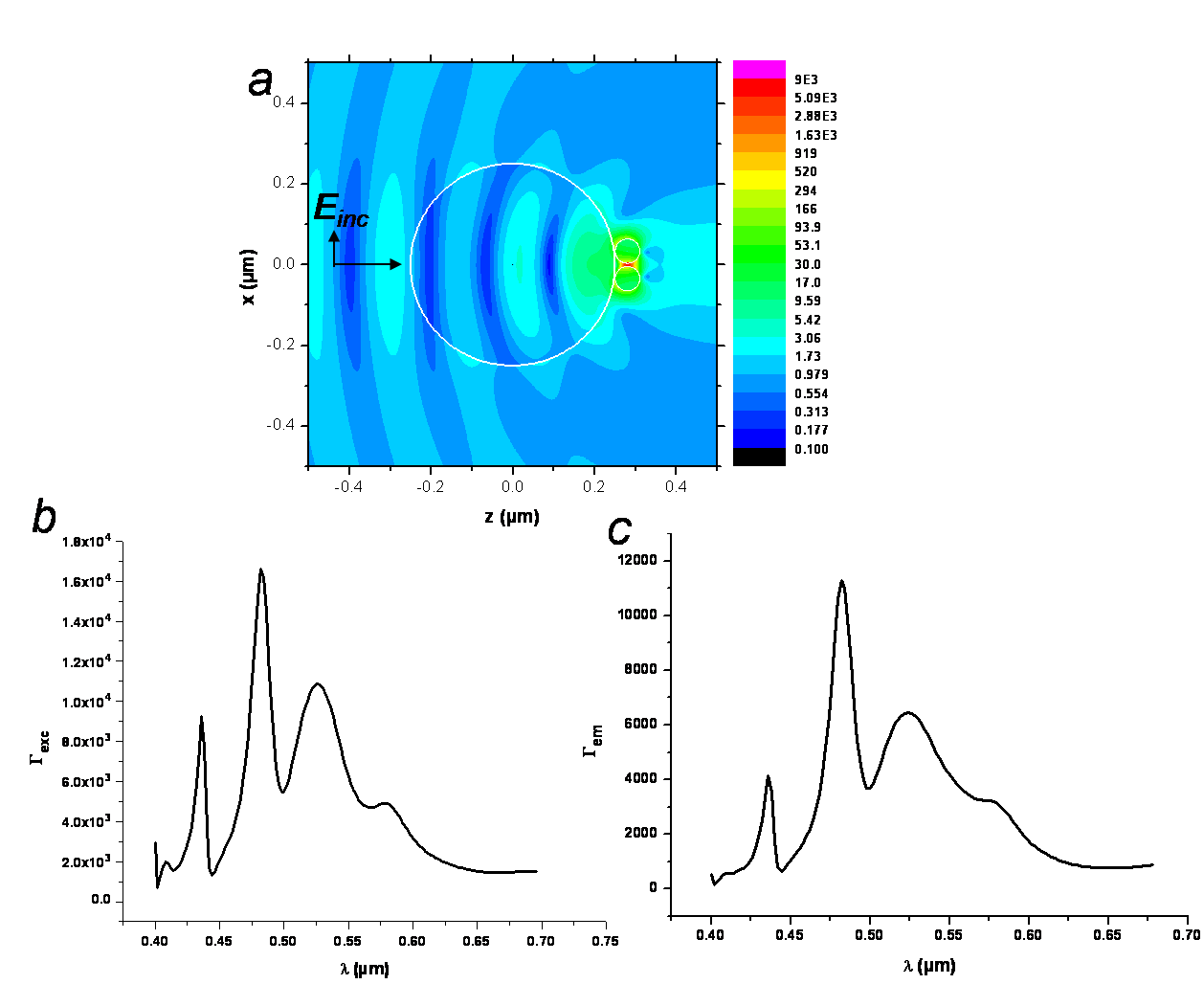}% Here is how to import EPS art
\caption{a) Colored map of the electric field intensity enhancement at $\lambda$~=~500~nm, b) excitation rate enhancement, c) emission rate enhancement as a function of the wavelength.}
\label{fig:emission}
\end{figure}

For the sake of completeness, we investigate the fluorescence enhancement of a molecule located in the vicinity of the nanoantenna. The fluorescence rate enhancement $\Gamma_{\rm em}$ can be defined as the product of the excitation rate $\Gamma_{\rm exc}$ and the quantum efficiency $\eta$ \cite{Quench_novotny, Colas_fluo}. The excitation rate is defined as ${|n_p.E(r_p)|}^2/{|n_p.E_0(r_p)|}^2$ where $n_p$ is the dipole direction and $E(r_p)$ and $E_0(r_p)$ are the local electric field at the dipole location $r_p$ respectively with and without the antenna. The enhancement of the electric field intensity is displayed in \ref{fig:emission}$a$ when the antenna is illuminated at $\lambda$~=~500~nm by a plane wave propagating along the $z$-axis. One observes a huge enhancement of light intensity in a nanometer sized volume delimited by the metallic spheres due to the combination of light focusing by the dielectric spheres with the excitation of coupled plasmons in the metallic dimer. This result was expected from \ref{fig:results}$d$ and reciprocity \cite{reciproc_greffet, spect_novotny} between the excitation and the radiative decay rates. Consequently, this antenna presents all the required properties for single molecular detection since both $\Gamma_{\rm exc}$, and $\eta$ are highly enhanced and concurrent with a high directivity. This supposition is confirmed by a calculation of the fluorescence rate enhancement displayed in \ref{fig:emission}$c$ as a function of the excitation wavelength $\lambda$, where the excitation rates and quantum efficiencies are calculated with a wavelength shift of 20~nm, representative of common fluorophores, and Alexa dye in particular. One observes that fluorescence enhancements as high as $10^4$ are achieved.

This study demonstrates that appropriately designed metallo-dielectric systems can serve as  compact, highly directive and ultra radiative antennas. Let us emphasize that contrary to fully metallic antennas, the high directivity of this antenna does not result from a plasmonic effect, and that it is efficient over a wide range of frequencies. In consequence, the high directivity does compromise the high radiative decay rate enhancement offered by two coupled metallic particles and it is possible to exploit whispering gallery modes to further enhance the radiative decay rates. This work paves the way towards the design of compact, simple and highly efficient optical antennas.

\providecommand*{\mcitethebibliography}{\thebibliography}
\csname @ifundefined\endcsname{endmcitethebibliography}
{\let\endmcitethebibliography\endthebibliography}{}

\end{document}